\documentclass{article}

\usepackage{arxiv}

\usepackage[utf8]{inputenc} 
\usepackage[T1]{fontenc}    
\usepackage{hyperref}       
\usepackage{url}            
\usepackage{booktabs}       
\usepackage{amsfonts}       
\usepackage{nicefrac}       
\usepackage{microtype}      
\usepackage{pifont}          

\usepackage[skip=14pt]{caption} 

\usepackage[pscoord]{eso-pic}
\newcommand{\placetextbox}[3]{
  \setbox0=\hbox{#3}
  \AddToShipoutPictureFG*{
    \put(\LenToUnit{#1\paperwidth},\LenToUnit{#2\paperheight}){\vtop{{\null}\makebox[0pt][c]{#3}}}%
  }%
}%

\usepackage{titlesec}
\makeatletter
\newcommand{\setappendix}{Appendix~\thesection:~}
\newcommand{\setsection}{\thesection~}
\titleformat{\section}{\bfseries\large}{%
  \ifnum\pdfstrcmp{\@currenvir}{appendices}=0
    \setappendix
  \else
    \setsection
  \fi}{0em}{}
\makeatother
\usepackage{appendix}

\usepackage{pslatex}
\usepackage[english]{babel}
\usepackage{multirow}
\usepackage{color}
\usepackage{cite}
\usepackage{graphicx}
\usepackage[hyphenbreaks]{breakurl} 
\usepackage{epstopdf}
\usepackage{makecell}
\usepackage{fixltx2e}

\usepackage{tabularx}

\usepackage{psfrag}

\usepackage{float}

\usepackage{xspace}
\newcommand{\MPEGH}{\mbox{MPEG-H Audio}\xspace}

\title{MPEG-H Audio for Improving Accessibility in Broadcasting and Streaming}

\date{\vspace{-3ex}} 

\author{
  Christian Simon\\
  Fraunhofer Institute for Integrated Circuits IIS\\
  Erlangen, Germany\\
  \And
  Matteo Torcoli\\
  Fraunhofer Institute for Integrated Circuits IIS\\
  Erlangen, Germany\\
  \And
  Jouni Paulus\\
  Fraunhofer Institute for Integrated Circuits IIS\\
  and\\
  International Audio Laboratories Erlangen\thanks{A joint institution of Universit{\"a}t Erlangen-N{\"u}rnberg and Fraunhofer IIS.}\\
  Erlangen, Germany
}

\begin{document}
\maketitle

\begin{abstract}
Broadcasting and streaming services still suffer from various levels of accessibility barriers for a significant portion of the population, limiting the access to information and culture, and in the most severe cases limiting the empowerment of people.
This paper provides a brief overview of some of the most common accessibility barriers encountered. It then gives a short introduction to object-based audio (OBA) production and transport, focusing on the aspects relevant for lowering accessibility barriers.
\MPEGH is used as a concrete example of an OBA system already deployed.
Two example cases (dialog enhancement and audio description) are used to demonstrate in detail the simplicity of producing MPEG-H Audio content providing improved accessibility. Several other possibilities are outlined briefly.
We show that using OBA for broadcasting and streaming content allows offering several accessibility features in a flexible manner, requiring only small changes to the existing production workflow, assuming the receiver supports the functionality.
\end{abstract}


\placetextbox{0.5}{0.06}{\fbox{\parbox{\dimexpr\linewidth-2\fboxsep-2\fboxrule\relax}{\centering The copyright of the this document is with the authors and their institutions.\\ A non-exclusive and irrevocable license to distribute the document was granted to arXiv.org.}}}%

\section{Introduction}
\label{sec:intro}
The delivery of audio-visual content to a broad audience has a hundred-year story, deeply woven together with technological innovations. 
The first radio was built at the end of the 19th century by Guglielmo Marconi and the first AM radio programs were broadcast in 1920 in the United States. 
Since then, innovations in broadcasting have appeared with an increasing pace: television (TV), FM, satellite, cable, color TV, portable devices, digitalization, the Internet, high definition (HD), streaming, smart devices, Ultra-HD TV.
These services constitute the main medium through which information and culture can reach people and contribute to their empowerment.  

However, for various reasons, today's broadcasting and streaming often have accessibility barriers for a significant portion of the population, creating a disabling environment. 
This paper gives an overview of these accessibility barriers~(Sec.~\ref{sec:issues}). 
Then, the main features of \MPEGH, an audio system supporting Object-Based Audio (OBA), are described (Sec.~\ref{sec:mpegh}). 
These can offer more accessible content (Sec.~\ref{sec:accessibilityOBA}). 
Finally, conclusions are given (Sec.~\ref{sec:conclusions}).

\section{Accessibility barriers in today's broadcasting and streaming}
\label{sec:issues}
The World Health Organization estimates that over a billion people (about 15\% of the world's population) have some sort of a disability~\cite{WHO:2018}. 
For these people, today's broadcasting and streaming may be in some way inaccessible.
This is because the audio-visual content is predominantly produced and delivered as a one-size-fits-all product, which cannot satisfy the diverse needs of a heterogeneous audience.
Some of the main accessibility issues related to broadcasting and streaming are reviewed in the following and summarized in Table~\ref{tab:overview}.
For a more extensive analysis of these issues the reader is referred to~\cite{EBU-04, Looms-11, EBU-16,ITUR-18}.
The technology described in this paper is framed within the inclusive and social models of media accessibility. 
Nevertheless, we use a medical model for some of the described barriers, because this facilitates the reference to the technical accessibility features mentioned in Sec.~\ref{sec:accessibilityOBA}.

\subsection{Awareness and budget}
Inadequate awareness and limited budget are the main causes for the encountered barriers.
On the one hand, it can be hard for decision-makers to gather correct information and statistics on accessibility issues, despite many associations being active in raising awareness on this.
On the other hand, the needed budget for addressing the accessibility problems is often not available, even if the problem is well-understood and a technological solution is available. 

\subsection{Age-related factors}
Broadcasting and streaming target an audience of any age, from children to elderly, all of which having very different needs. 
Age can be one of the causes of the accessibility problems described in the following subsections, e.g., the language complexity level can be too high for the children, or the audio level of the dialog compared to the background level can be too low for the elderly.
Hearing is an important age-related factor, often degrading with age; estimates are that one third of people over 65 years of age are affected by disabling hearing loss~\cite{WHO:2019}.
This is of great interest for broadcasters, because the average age of the audience is already high and keeps on increasing, e.g., for BBC One, the average watcher was 56 year old in 2011, 59 years in 2014, and 61 years in 2017~\cite{BBCTrust:2017}. 
	
\subsection{Hearing}
Struggling to follow dialogs and audio cues is a common situation for people with hearing loss. 
This also happens for people without hearing impairments but consuming the content in a loud environment and/or using a low quality reproduction system (e.g., low-quality headphones connected to a smartphone while in a bus). 
One of the most common complaints to broadcasters is about the low intelligibility of the speech due to the loud background music and noise~\cite{armstrong:2016}, preventing the audience from understanding and enjoying the content. 
Missing non-speech audio-only cues with high narrative importance also affects the understanding of the program~\cite{Ward17-INTERSPEECH, Shirley17-JAES, Ward18-AES}. 
Users would benefit from being able to personalize the relative level of the main audio elements composing the audio mix.

\subsection{Language}
The vast majority of today's TV programs are available in only one language. 
However, the language spoken in the content constitutes a barrier for people who do not understand it. 
This is especially problematic for countries with more than one official language or with a number of languages spoken regionally 
or by immigrant groups (e.g., the Spanish language in the USA). 
Providing multiple languages for the same content (or translating captioning) can be a vehicle of social integration and promote social cohesion. 
Another language-related problem is that the complexity level of the spoken language exceeds the capabilities of some people, e.g., when they are just learning the language (e.g., children and non-native speakers) or due to cognitive disabilities or fatigue. 
These people could benefit from an alternative version where a simplified vocabulary and/or a lower pace are used (an experimental speech rate converter is presented in~\cite{ITUR-18}).
Even when non-native speakers are fluent in the foreign language, they can benefit from a level of the dialog that is higher than the one suitable for native speakers \cite{Florentine85-JASA}.
Finally, a very small portion of today's content comes with sign language, which is the primary language of a relevant portion of people with severe hearing loss.  
	
\subsection{Sight}
It has been estimated that 1.3 billion people live with some form of vision impairment and that 36 million people are blind~\cite{WHO:2018blind}. 
People with sight loss also follow a lot of television, or would certainly like to do so~\cite{Looms-11}. 
In order to fully understand and enjoy the content, audio description (AD) is needed to understand the information otherwise transported via visual cues. 
Moreover, in the case of subtitles translating a dialog from a foreign language, these have to be added to the audio representation as spoken subtitles\footnote{Spoken subtitles (also known as audio subtitling) consist of a voice reading the given subtitles. For economic reasons, this is often synthesized speech. In the Nordic Countries of Europe, almost all programs are offered with spoken subtitles \cite{EBU-16}.}.
Only a small portion of today's content comes with AD due to the additional production costs involved.
Moreover, this is usually broadcast on a dedicated audio channel, which is not fully inclusive and its activation may introduce additional technical difficulties.
In the UK, the Ofcom prescribes that at least 10\% of the broadcast programs shall be available with AD~\cite{Ofcom:2017}.

The recent advances in speech synthesis, e.g.,~\cite{van2016wavenet}, and a better understanding of speech mixing~\cite{torcoli:2019} give the tools for automatically and more economically creating AD from a script.
On the other hand, creating content with high-quality AD is a topic of current research, e.g.,~\cite{lopez:2009,portillo:2018}.

\subsection{Literacy}
Using on-screen text (e.g., subtitles) assumes that the viewer is a proficient reader, which is not always the case. Subtitles often require the viewer to be able to read very quickly. Estimates are that even in countries with high literacy levels, as many as 10-20\% will not be able to follow on-screen texts \cite{Looms-11}. 
This group includes also the people not familiar with the alphabet used in the subtitles, even if they would be able to read quicker in another alphabet.
As a solution, the language can be condensed so to bring the required reading speed down to acceptable levels. Alternatively, spoken subtitles can be provided.
If accessible, broadcasting and streaming can have an educational role and help illiteracy eradication.
	
\subsection{Cognition}
The audio-visual content may be difficult to enjoy or understand also because of a number of permanent or temporary cognitive disabilities. 
This is a very vast and complex field and its analysis is beyond the scope of this paper. 
Just a few examples are mentioned. 
These disabilities can include cognitive fatigue, aphasia (a.k.a.~''word blindness”), and dementia. 
Also in these cases, aforementioned accessibility features such as personalized relative audio level, captioning, AD, spoken subtitles, and alternative simplified language could be of assistance.
People on the autistic spectrum may not be able to identify or correctly interpret social or emotional visual cues and would benefit from AD explicitly noting these.

\subsection{Mobility and dexterity}
An important barrier can arise while setting up and operating a modern receiver. 
This is especially the case for persons with manual dexterity impairment, but it can also be difficult for people with cognitive disabilities or low literacy level. 
Voice-based user interfaces and appropriately designed presets can be of help in some cases. This has mostly to do with the receiver interface design, which is beyond the scope of this paper. 

We can note that many of the mentioned accessibility problems are in some way related to the audio modality and audio content reproduction. 
This suggests that an appropriate audio transport solution for the broadcasting and streaming allows addressing many of the named issues making the content more accessible. 

\section{Object-Based Audio with \MPEGH}
\label{sec:mpegh}
Object-Based Audio (OBA) is a broad term that refers to the production and delivery of sound based on \emph{audio objects}. 
In this context, an audio object represents a \emph{component of the audio mix} delivered separately to the receiver and to which \emph{metadata} have been added~\cite{simon2018}.
Together the audio and the metadata information are referred to as \emph{audio scene} and encoded into one stream~\cite{MPEG-H_Metadata}.
On the decoder side, separate objects can be made available and controlled by the metadata information. 
Hence, OBA supports three main innovative features. These are:
\begin{enumerate}
\item \textbf{Immersive sound}, i.e., the possibility of creating three-dimensional sound scenes immersing the listener.
\item \textbf{Universal delivery}, ensuring optimized reproduction across different classes of playback devices, e.g., over different loudspeaker setups (a soundbar, a TV, or a mobile device), all  from one single audio stream.
\item \textbf{Advanced user-interactivity}, enabling the user to personalize the final audio mix to their needs and taste, e.g., by personalizing the level and the position of the dialog. The interactivity can be controlled, e.g., via a remote control or a voice user interface.
\end{enumerate}

\begin{figure}[t] 
	\centering
	\includegraphics[width=0.8\columnwidth]{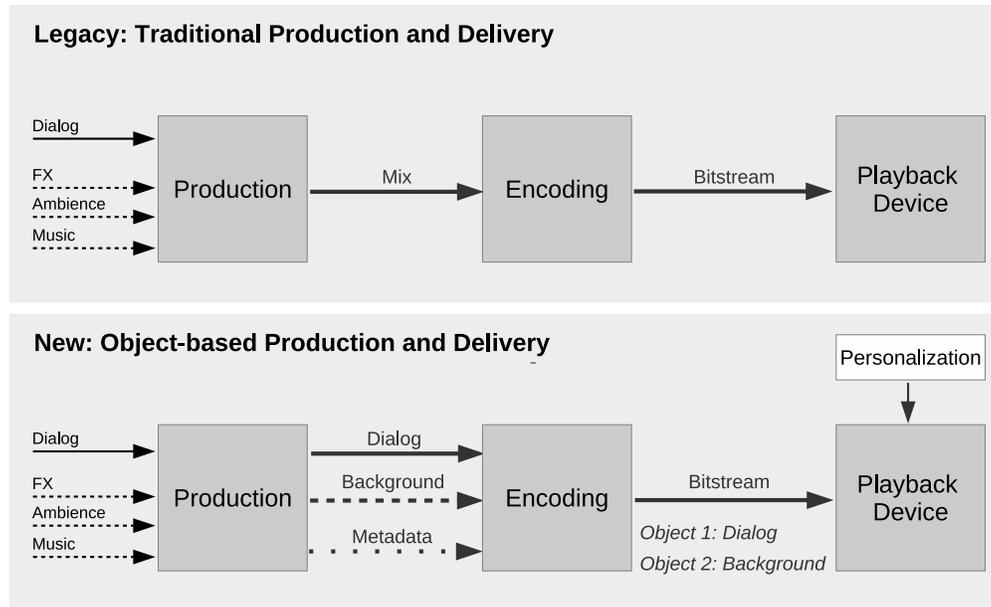}
	\caption{While in a legacy workflow only one, immutable mix is produced and delivered, an object-based workflow enables personalization features in the playback device by delivering the components of the audio scene with their attached metadata.}
	\label{fig:Legacy_vs_OBA}
\end{figure}

These features are revolutionary with respect to traditional (or legacy) audio, where immutable audio mixes are produced and delivered to the consumer, see Fig.~\ref{fig:Legacy_vs_OBA}.
These innovative features (and in particular the user-interactivity) can significantly improve the accessibility of broadcasting and streaming, as discussed in Sec.~\ref{sec:accessibilityOBA} and summarized in Table~\ref{tab:overview}.
This paper focuses on the OBA system of \MPEGH~\cite{Herre14-JAES}.

The rest of this section briefly describes the most important aspects of \MPEGH metadata (Sec.~\ref{sec:metadata}), and summarize the basics of \emph{authoring} metadata (Sec.~\ref{sec:authoring}), encoding, transmission, and decoding the content (Sec.~\ref{sec:encodingAndCo}).

\subsection{Metadata}
\label{sec:metadata}
Metadata are the second main component (in addition to the audio itself) in OBA allowing the novel functionality.
It carries information describing the audio, e.g., scene presets, component and overall loudness, dynamic range control data, spatial location and possible movement of the audio objects, to name a few. 
Appendix~\ref{app:metadata} describes more in details examples of metadata.
The metadata are fundamental for enabling and controlling the features offered by OBA. 

\subsection{Authoring}
\label{sec:authoring}
\emph{Authoring} is the process of creating the object-based audio representation from the components by associating the audio with the appropriate metadata.
This defines the relationships among the audio objects in the audio scene and how the user can interact with them.
The objects and presets are labeled and the allowed interactivity ranges are defined at this stage.
In addition to the creation of the content, it is equally important to monitor all presets, interactivity features, and downmixes to common lower-order layouts, in order to make sure the additional functionality works in the desired way.
The authoring process ends with the export of the audio and its associated metadata.
The current \MPEGH authoring tools can support different types of metadata export: 
\begin{enumerate}
	\item As part of an  \textbf{ADM} file~\cite{ITU-ADM}.
	\item As \textbf{Control Track}, modulated in an audio channel, comparable with an audio timecode track~\cite{Bleidt-17}.
\end{enumerate}

The latter, combined with the audio components of the OBA scene in a multichannel audio file is called MPEG-H Production Format (MPF).
This proprietary format has the advantage of being transmittable over regular audio connections like SDI.

There are \MPEGH authoring tools available for both live and post-production applications.
An example is the MPEG-H Authoring Plugin, which is a freely-available tool compatible with most important digital audio workstations\footnote{\url{https://www.iis.fraunhofer.de/de/ff/amm/dl/software/mhapi.html}}.

\subsection{Encoding, transporting, and decoding}
\label{sec:encodingAndCo}
After the authoring process, the material is ready for encoding.
This can be accomplished simply by feeding the created MPF file together with the associated video into a live or an offline encoder.
Such programs and devices are already on the market and in daily use in broadcast workflows.
The created file or stream can be saved, streamed, or broadcast similar to legacy content. 
The decoding and playback of the audio stream takes place in an \emph{MPEG-H enabled end-user device}, taking into account the user's personalization and rendering settings.
The \MPEGH codec is already standardized in several international broadcast and streaming standards, e.g, ATSC~3.0~\cite{Bleidt-17}, 
DVB-MPEG/UHD~\cite{DVB_TS_101_154}, DVB-DASH~\cite{DVB_TS_103_285}, TTA~\cite{TTA16}, HbbTV~\cite{HbbTV18}, to name a few. 

\section{Accessibility with Object-Based Audio}
\label{sec:accessibilityOBA}
As introduced in Sec.\ref{sec:issues}, a very broad range of accessibility barriers are encountered in today's broadcasting and streaming. 
In the following, the potential solutions offered by OBA and the introduced benefits are discussed. These are also summarized in Table~\ref{tab:overview}.
Practical descriptions on how to produce MPEG-H content are given in Appendix~\ref{workflows}.

\subsection{Addressing accessibility barriers with OBA}
\label{sec:addressing}
A key  accessibility feature is the \textbf{user interactivity} offered by OBA. 
Thanks to this, each person can personalize the relative level of the dialog and of other important audio cues. The output overall loudness remains homogeneous thanks to the real-time loudness adaptation in the MPEG-H decoder. 
One can also configure the playback device to automatically select (if available) the ``Dialog+'' preset, i.e., a version of the audio mix where the dialog has a higher relative level. 
In such a way, minimal interaction with the receiver is needed. 
Higher levels of the dialog are especially useful when the user's hearing is hindered, for example, due to age-related hearing loss.
Moreover, this can be beneficial for fluent non-native speakers.

The MPEG-H \textbf{Dynamic Range Control (DRC)} adapts the dynamic range and the level of the output signal to the individual playback device and situation. E.g., a compressed dynamic range is used for playback on a smartphone to address the limitations of the small internal speakers or the masking introduced by a noisy environment. E.g., this can be helpful for the intelligibility of the dialog, as its quieter parts are made louder.

The transmission of \textbf{multiple languages or alternative versions} of one language can lower or eradicate problems related to the language or its complexity level.
One of the alternative versions can include audio description or spoken subtitles, improving the accessibility for visually impaired people. 
Spoken subtitles can also help people with limited literacy abilities.
Each language (version) can be transmitted as an object to be overlapped on top of the same background, which can be immersive. Hence, e.g., the watchers of the AD version can enjoy the same immersive sound quality as all other users. This is often not the case in today's broadcasting, where the AD mix is frequently available only as stereo, even if the content without AD features an immersive mix, in order to spare transmission bandwidth.

One asset of OBA that could address users with mobility or cognitive disabilities is the transmission of all accessibility features in \textbf{one stream} and the automatic selection of the \textbf{preferred preset kind}, which has to be set only once for each device. This makes a repeated selection of the wanted content representation unnecessary. It has to be noted that the implementation of the user interface is outside the scope of OBA and it is left to the playback device manufacturers.

Barriers caused by limited budgets cannot be directly addressed by OBA. 
But there are different signal processing techniques, such as dialog separation~\cite{paulus:2019} and automatic mixing algorithms, which can speed up the production significantly and bring a benefit in the context of dialog enhancement and general accessible content creation.
These techniques can be easily connected with OBA. 
Mobility disabilities can not be addressed by OBA systems.

\begin{figure}[t] 
	\centering
	\includegraphics[width=1\columnwidth]{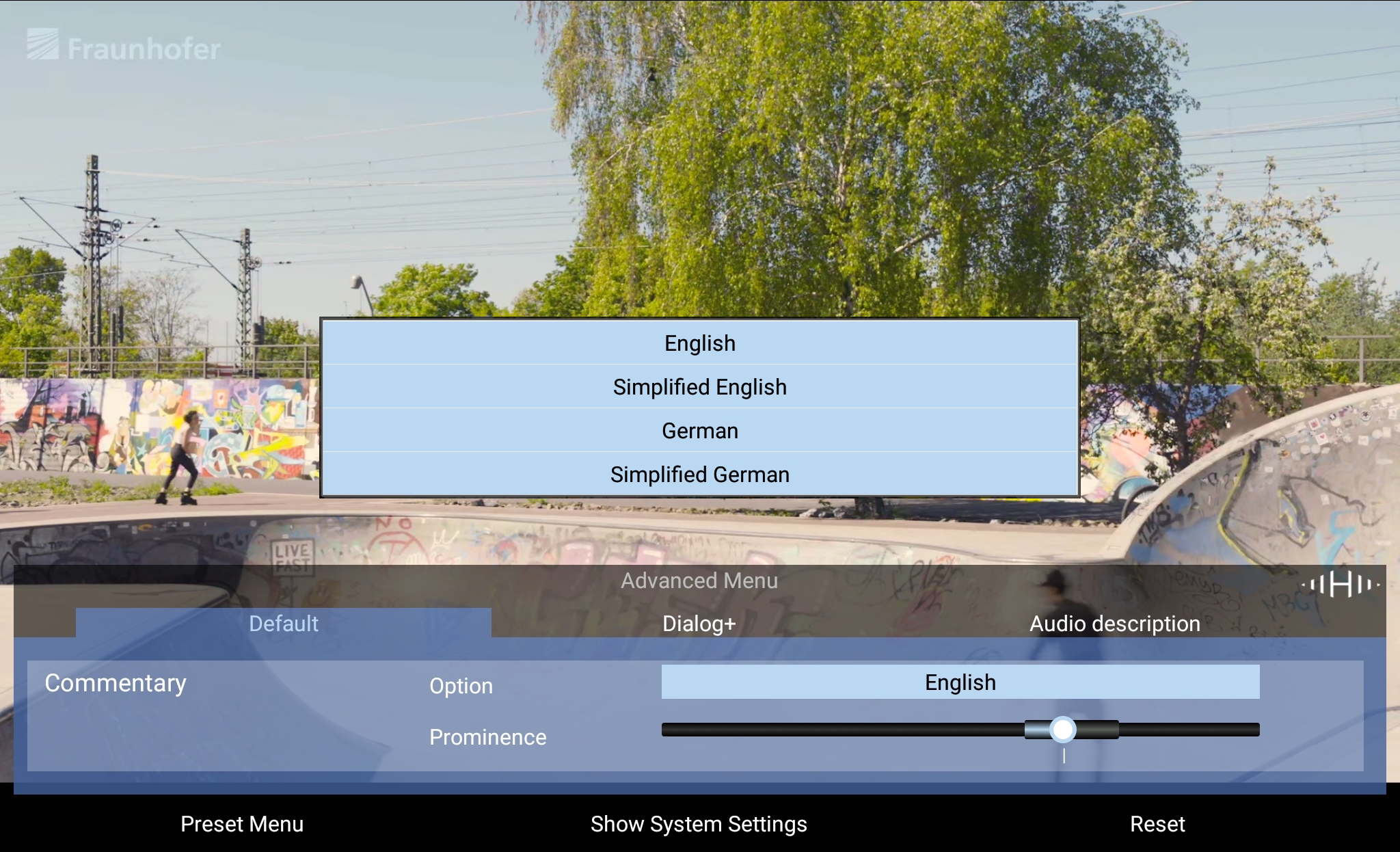}
	\caption{Screenshot of an accessible \MPEGH audio scene with three presets: The ``Default'' preset contains four different language versions including simplified language. The ``Dialog+'' preset provides an audio version with  better speech intelligibility. The ``Audio description'' preset is self-explanatory. The open dialog box allows selecting the audio language and version.}
	\label{fig:Big_audio_scene_screenshot}
\end{figure}

Fig.~\ref{fig:Big_audio_scene_screenshot} shows an exemplary screenshot of content authored with MPEG-H, providing a variety of the accessibility features named above.

\subsection{Benefits from using object-based audio}
\label{sec:oba_benefits}

Summarizing, making use of the OBA features provides immediate accessibility benefits for the end-user, such as:
\begin{itemize}
\item Better speech intelligibility is achieved by allowing gain interactivity of the dialog object.
\item Multiple language versions and AD are available within the same audio stream.
\item Accessible presets can be played back automatically if it is present in the stream.
\item Accessible presets are a part of the regular broadcast.
\item Personalization of audio object level and position is possible.
\item Users of versions with accessible audio can also enjoy multichannel mixes.
\end{itemize}
Moreover, also the broadcasters and the content-providers benefit from making use of the new features:
\begin{itemize}
\item Some of the regulations demanding accessibility features are fulfilled.
\item No need to make a compromise between creativity and accessibility.
\item One stream includes all versions of the content making data delivery easier and more inclusive.
\item Full control over the authoring process allows defining how the content may be modified in the playback.
\item Lower bandwidth may be achieved, as the AD track is only an additional mono track with the associated metadata instead of a full, possibly multichannel, mix.
\end{itemize}

\section{Conclusions}
\label{sec:conclusions}
In this paper we have presented an overview of some accessibility barriers present in today's broadcasting and streaming.
This was followed by an introduction to object-based audio (OBA), its main principles, features, and benefits for accessibility, focusing the description to the specific OBA system of \MPEGH.
As summarized in Table~\ref{tab:overview}, we have shown the potential of breaking many of the accessibility barriers by using features of MPEG-H.

In Appendix~\ref{app:metadata} we give a brief overview about MPEG-H metadata types relevant for accessibility.
In Appendix~\ref{workflows} we provide concrete workflow examples needed to use \MPEGH for producing content with user-selectable dialog-enhanced audio version or an AD audio track, both making use of the new possibilities offered by OBA.
All these examples show that it is possible to make broadcasting and streaming much more accessible if OBA (e.g., \MPEGH) is used for the transport of the content and the end-user has a receiver supporting the functionality.

\begin{table}[htbp]
\centering
\def\arraystretch{1.5}
\begin{tabular}{|p{20mm}|p{20mm}|p{25mm}|p{40mm}|p{36mm}|}
\hline

\textbf{Context} & \textbf{Barrier} & \textbf{Cause} & \textbf{Involves people}  & \textbf{OBA solution} \\ 
\hline
\hline

Hearing & Dialog is not understood & Dialog level is too low compared to background level or in general &
$\bullet$ Hard of hearing \newline 
$\bullet$ In noisy environment \newline 
$\bullet$ Using low-quality play-back system \newline 
$\bullet$ Using low overall level (late night) \newline
$\bullet$ Non-native speakers &  
\ding{51}  \textbf{Dialog level} can be \textbf{personalized} by the end-user.\newline
\ding{51}  \textbf{Dynamic range control (DRC)} adapts signal level and dynamic range.\\
\hline

Hearing & Audio-only cues of narrative importance are not understood &  Content relies on a single modality for essential information &
$\bullet$ Hard of hearing \newline 
$\bullet$ In noisy environment \newline 
$\bullet$ Using low-quality playback system \newline 
$\bullet$ Using low overall level (late night) &  
\ding{51}  \textbf{Audio object levels} can be \textbf{personalized} by the final user, e.g., emphasizing information with high \emph{narrative importance}.\newline
\ding{51}  \textbf{DRC} adapts signal level and dynamic range.\\ 
\hline

Language & Dialog is not understood & Spoken language is not understood  &
$\bullet$ Living in a foreign country \newline 
$\bullet$ Living in country with more than one language (officially or de-facto) &  
\ding{51}  \textbf{Multiple languages} can be carried in the same stream as objects efficiently.\\ 
\hline

Language complexity & Dialog is not understood & The level of the language is too difficult or the pace is too high &
$\bullet$ Non-native speakers \newline
$\bullet$ With cognitive disabilities \newline
$\bullet$ Children & 
\ding{51} Dialog with \textbf{simplified vocabulary or a lower pace} can be carried in the same stream as objects efficiently. \\ 
\hline

Sight & Visual cues are not understood & Visual information is partially or completely missed & 
$\bullet$ Blind or partially-sighted \newline
$\bullet$ Following the program as audio-only \newline 
$\bullet$ Autistic spectrum people &
\ding{51} \textbf{Audio description} can be carried in the same stream as an object, and its level and position can be personalized. \\ 
\hline

Literacy or sight & Subtitles (e.g., translating a foreign language) are not understood & Reading ability or comfort & 
$\bullet$ With cognitive disability\newline
$\bullet$ Children \newline
$\bullet$ Blind or partially-sighted \newline
$\bullet$ Following the program as audio-only  &
\ding{51} \textbf{Spoken subtitles} can be carried in the same stream as an object, and their level and position can be personalized. \\ 
\hline

Mobility or cognition & The receiver cannot be set up or operated & Difficulty in operating the user interface & 
$\bullet$ With limited mobility and dexterity  \newline
$\bullet$ With cognitive disability & 
This is \textbf{outside the scope} of OBA and left to the receiver manufacturers. Setting the preferred \textbf{preset kind} could help.\\
\hline

Budget & Accessibility feature is not available & Production costs of separated dialog, additional languages, AD, subtitling, sign language, etc.& 
$\bullet$ Everybody &  
Even if \textbf{outside the scope} of OBA, the combination of OBA transport with \textbf{dialog separation, speech synthesis, and automatic mixing} has been successfully tested. \\ 
\hline
\end{tabular}
\vspace{5pt}
\caption{\label{tab:overview}Overview on accessibility barriers in today's broadcasting and solutions enabled by object-based audio (OBA).}
\end{table}

\begin{appendices}
\newpage
\section{Metadata}
\label{app:metadata}

The \MPEGH metadata carry information regarding presets, loudness, DRC, 3D audio, and more, as described in the following.

\subsection{Presets and labels}
\label{Presets and labels}
Presets are the basic interaction mode of \MPEGH. 
They allow the content producer to create multiple representations (or versions) in one delivered audio scene, which then can be easily selected in the playback device. For example, ``Default mix'' and ``Dialog+'' could be two versions with different relative levels of the dialog object.
A preset can also contain different languages, which can be selected in the playback device.
A preset can be defined to be of a specific preset kind, e.g., audio description (AD), which can be automatically selected by the playback device, if the preset kind is available and the user has enabled this feature.
In this way, the users need to define their preferences only once, and the desired version of the content will be played back.

Labels allow naming the presets and audio components (groups of audio objects) individually for displaying the information on the user interface.
The labeling can also be multi-lingual and the correct language is then displayed according to the setting of the playback device.


\subsection{Loudness}
\label{Loudness}
Loudness information for every audio component and preset is measured and stored while authoring the audio scene.
This piece of information is used in the rendering stage for adapting the loudness of the various presets to a common value in order to prevent loudness jumps among presets.
The loudness adaption works in real-time to account also for the user interactivity, possibly changing the composition of the rendered audio scene.

\subsection{Gain and DRC}
\label{Gain and DRC}
Gain and dynamic range control (DRC) metadata provide gain information attached to a component or preset contained in the audio scene.
When a preset containing gain metadata is chosen, the gain information is applied to the corresponding audio components~\cite{MPEG-H_DRC}. 
For example, dynamic gain information can be used to lower the level of a regular film mix every time an AD voice-over is active.
Furthermore, if gain interactivity is allowed for a component, the level of this component can be increased or decreased by the user within the allowed interactivity range, e.g., adjusting the level of the voice-over relative to the rest of the mix.
While doing this the \MPEGH decoder adjusts the overall level in order to avoid loudness jumps.
The DRC metadata allow adapting the dynamic range of the output signal to the individual playback device and situation, e.g., using a smaller dynamic range for playback on a smartphone to address the limitations of the small internal speakers or the noisy environment, or using a wider dynamic range on AVR playback.

\subsection{Position, immersive sound, and downmixes}
\label{Position}
Position metadata control the location of the audio objects in the 3D audio scene. Similarly to the gain metadata, position metadata define the allowed range for interactivity.
For example, with position interactivity the AD voice-over can be placed at a rear speaker of a surround speaker setup, making the AD speaker ``whispers into the ear'' of the listener, while the full film mix stays at its normal position.

In OBA, the audio scene is authored and transmitted in the highest channel order for which the content is intended for, and the reproduction of this scene on a device with fewer output channels requires downmixing the audio.
The advantage of an OBA system is that the downmixes do not need to be done channel-based during the authoring and transmitted individually, but they are taken care by the rendering engine in the playback device. 

\section{Workflows for producing MPEG-H content}
\label{workflows}
This appendix describes two example workflows for producing MPEG-H content in practice. 
The first workflow is for improving speech intelligibility, while the second one is for adding AD. The workflows can be combined, but they are presented separately for the sake of clarity. With very similar workflows, one could create multiple language versions of the same content, provide a simplified dialog track, add spoken subtitles, etc.

The two examples will show that using an object-based production workflow to make the content more accessible is straightforward. Often the modification with respect to the traditional production workflow simply consists in omitting the creation of the final full mix. Instead, the component signals are exported and appropriate metadata are then associated to those components in the authoring process.

\subsection{How to create an MPEG-H scene with two presets (Default and Dialog+)}
\label{Speech intelligibility}
\begin{figure}[t] 
	\centering
	\includegraphics[width=1\columnwidth]{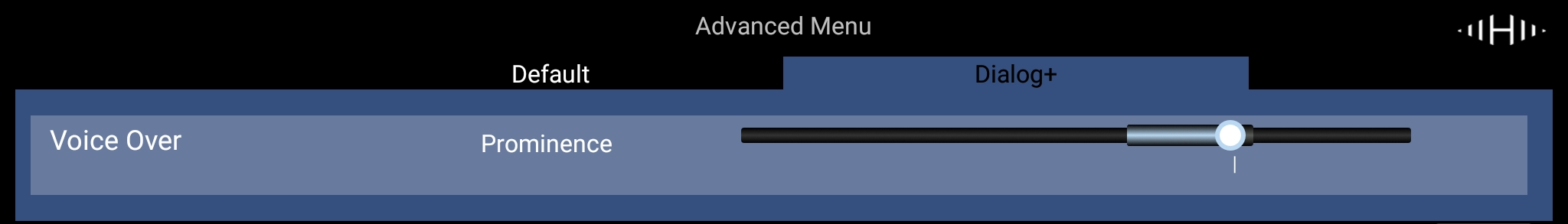}
	\caption{A screenshot of an exemplary \MPEGH user interface shows the advanced menu of the audio scene described in Sec.~\ref{Speech intelligibility}.
	The lighter part of the ``Prominence" bar denotes the allowed interactivity gain range of the ``Voice-over'' object, and the white slider denotes the current setting.}
	\label{fig:GUI_Dialog+}
\end{figure}

The next two sections describe hands-on examples of content creation with OBA.
We start by describing how to create an MPEG-H audio scene with two presets with differently authored voice-over mixes: one for ``regular film mix'' and one for enhanced dialog intelligibility.
Gain interactivity for the voice-over object is allowed in both presets, allowing the user to adjust the level of the voice-over compared to the rest of the mix. 
Additionally, the preset with the enhanced dialog intelligibility should be selected automatically if this functionality has been activated in the playback device settings. The content-creation follows the steps of:
\begin{itemize}
\item The production starts similar to a legacy production considering mixing the voice-over:
The background audio, often referred to as \emph{international tape} (IT), gets attenuated in an aesthetically pleasing way every time the voice-over is active~\cite{torcoli:2019}.
However, the voice-over and the dipped\footnote{{``Dipped" is a common term for a mix which is attenuated at positions in which a voice-over is active. }} IT are not mixed together to create a regular, channel-based audio mix, but are kept as separate tracks, and considered as audio objects.
\item An MPEG-H authoring tool is used to create two audio scene presets consisting of the audio tracks and the associated metadata.
Both presets contain the voice-over and the dipped IT audio objects. 
\item The presets are labeled, in this example ``Default mix" and ``Dialog+", and assigned with a correct \emph{preset kind} metadata as described in Sec.~\ref{Presets and labels}, e.g., ``High-quality loudspeakers'' and ``Hearing impaired''.
Defining the correct preset kind allows the decoder to recognize the second preset as a version with enhanced dialog, and to automatically select it for the playback, assuming it is desired by the user settings.
\item The gain metadata are defined for both presets, allowing gain interactivity by some amount, e.g., $\pm$9\,dB. 
Additionally, the preset ``Dialog+" is assigned with a static gain offset of, e.g., +6\,dB to achieve a globally increased voice-over level and thus also improved speech intelligibility.
\item For quality control, the entire audio scene is monitored with the authoring tool, the interactivity features and possible downmixes are tested, and the settings are adjusted in the case of undesirable behavior.
\item Finally, the audio and metadata are exported as an ADM or MPF file and can be provided as an input to the actual encoder.
\end{itemize}
On the playback side in the decoder, the user-selected preset and possible interactivity inputs control the rendering of audio scene.
Figure \ref{fig:GUI_Dialog+} shows an exemplary user interface that results from  the workflow described above.

There is a significant amount of legacy content in which the dialog is not available as a separate track and still the functionality of dialog enhancement would be beneficial for the end-user.
In such a case, using source separation methods for splitting the mixture signal into a dialog and background tracks can be applied, as described, e.g., in~\cite{paulus:2019}.
After obtaining the separated component signals, the authoring process is similar to the one described above for separately-available component signals.

\subsection{How to create an \MPEGH scene with AD preset}
\label{Audio description}
The second example describes creating an audio scene with an AD preset in addition to the regular film mix.
The main feature of the AD preset is that the additional AD voice-over component has both gain and position interactivity enabled, and this preset should be automatically selected if the user-settings request AD playback.
The creation of this content follows these steps:
\begin{itemize}
\item The audio description mix is done similar as in a legacy production.
In other words, the full film mix gets attenuated in an aesthetically pleasing way every time the AD voice-over is active~\cite{torcoli:2019}.
Opposed to a legacy production, when the mix is finished, the voice-over and film mix are not mixed together to create a regular, channel-based AD mix, but they are kept as separate tracks. Instead, the created volume automation (the gain modification curve to be applied on the film mix) of the Digital Audio Workstation (DAW) is exported and used in the following authoring process.
\item An MPEG-H authoring tool is used to create an audio scene with two presets consisting of all audio tracks and the associated metadata, as described in Sec.~\ref{sec:authoring}.
\item Both presets are assigned with user-friendly labels, e.g, ``Default" and ``Audio description", and they are assigned with the correct \emph{preset kind} metadata, as described in Sec.~\ref{Presets and labels}, e.g., ``High-quality loudspeakers'' and ``Audio description''. In this example, the ``Default" preset contains only the original audio of the regular film mix, and the ``Audio description" preset consists of the audio of the regular film mix and the AD voice-over.
\item The volume automation curve from the DAW is converted to dynamic gain metadata and attached to the ``Audio description'' preset.
This gain will be activated when the preset is chosen, and it lowers the level of the regular film mix under the voice-over track. Opposed to the traditional production flow, the gain is applied first during playback in the end-user's device and not during the mixing process.
\item Gain and position interactivity of the AD voice-over is enabled, e.g., allowing adjustment of the AD level by $\pm$6\,dB, and the position by $\pm$180\,degrees in the horizontal plane and 0..$+$30\,degrees in the vertical axis.
\item The quality control and actual encoding take place similar to producing content with enhanced speech intelligibility (Sec.~\ref{Speech intelligibility}).
\end{itemize}

\begin{figure}[t] 
	\centering
	\includegraphics[width=1\columnwidth]{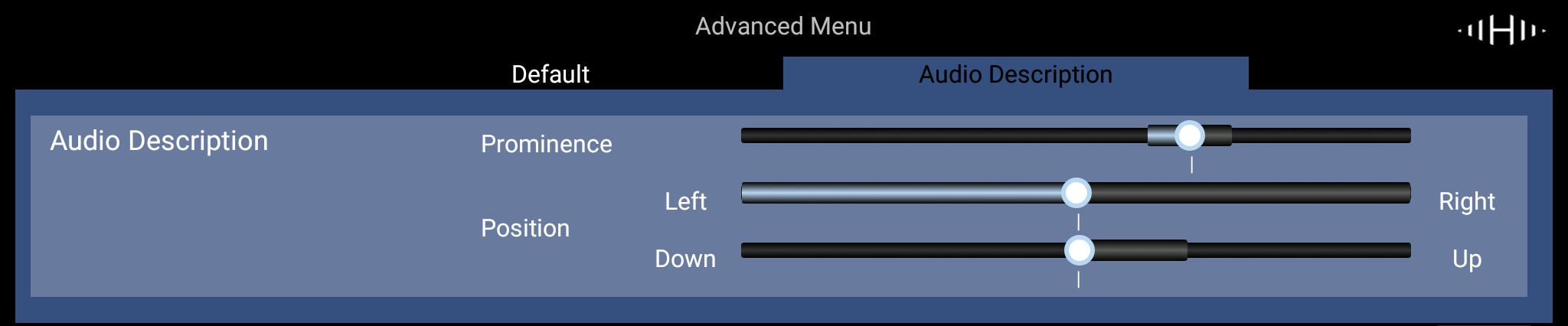}
	\caption{A screenshot of an exemplary \MPEGH user interface shows the advanced menu of the audio scene described in Sec.~\ref{Audio description}.
		The lighter part of the "Prominence" bar marks the allowed interactivity range of the ``Audio Description object'', and the white slider the current setting. 
		The other two sliders control the horizontal and vertical position of the AD object in their defined ranges.}
	\label{fig:GUI_AD}
\end{figure}

Figure \ref{fig:GUI_AD} shows an exemplary user interface that results from  the workflow described above. 
The interface is designed in two layers, one only showing the presets, and an advanced menu providing more elaborate setting of the single objects, if allowed by the content-provider.

\end{appendices}

\bibliographystyle{ieeetr} %
\bibliography{references}

\end{document}